\documentclass[sigconf]{acmart}

\AtBeginDocument{%
  \providecommand\BibTeX{{%
    \normalfont B\kern-0.5em{\scshape i\kern-0.25em b}\kern-0.8em\TeX}}}

\copyrightyear{2024}
\acmYear{2024}
\setcopyright{acmlicensed}\acmConference[KDD '24]{Proceedings of the 30th ACM SIGKDD Conference on Knowledge Discovery and Data Mining}{August 25--29, 2024}{Barcelona, Spain}
\acmBooktitle{Proceedings of the 30th ACM SIGKDD Conference on Knowledge Discovery and Data Mining (KDD '24), August 25--29, 2024, Barcelona, Spain}
\acmDOI{10.1145/3637528.3671569}
\acmISBN{979-8-4007-0490-1/24/08}

\usepackage{algorithm}
\usepackage{algorithmic}
\usepackage{pdfpages}
\usepackage{multirow}
\usepackage[export]{adjustbox} 
\usepackage{bm}
\usepackage{amsmath}
\usepackage{array}
\usepackage{booktabs}
\usepackage{footmisc}

\usepackage{amsthm}
\usepackage{graphicx}
\usepackage{enumitem}

\theoremstyle{definition}
\newtheorem{definition}{Definition}
 
\newtheorem{corollary}{Corollary}

\usepackage{subfigure}
\usepackage{subcaption}
\usepackage{hyperref}
\usepackage{url}
\usepackage{color}
\usepackage{xcolor}
\usepackage{colortbl}
\usepackage{xspace}

\newcommand{\pog}{POG\xspace}
\newcommand{\model}{POGCN\xspace}

\usepackage{amsmath,amsfonts,bm}

\def\eqref#1{equation~\ref{#1}}

\def\1{\bm{1}}

\def\ve{{\bm{e}}}

\def\mA{{\bm{A}}}
\def\mB{{\bm{B}}}

\def\mD{{\bm{D}}}
\def\mE{{\bm{E}}}

\def\mR{{\bm{R}}}

\DeclareMathAlphabet{\mathsfit}{\encodingdefault}{\sfdefault}{m}{sl}
\SetMathAlphabet{\mathsfit}{bold}{\encodingdefault}{\sfdefault}{bx}{n}

\def\gB{{\mathcal{B}}}
\def\gC{{\mathcal{C}}}
\def\gD{{\mathcal{D}}}

\def\gG{{\mathcal{G}}}

\def\gI{{\mathcal{I}}}

\def\gL{{\mathcal{L}}}

\def\gR{{\mathcal{R}}}

\def\gU{{\mathcal{U}}}

\def\gY{{\mathcal{Y}}}

\begin{document}

\title{Multi-Behavior Collaborative Filtering with Partial Order Graph Convolutional Networks}
\author{Yijie Zhang}
\authornote{All authors contributed equally to this research.}
\affiliation{%
  \institution{Jinan University}
  \city{Guangzhou}
  \country{China}
}
\email{wingszhangyijie@gmail.com}

\author{Yuanchen Bei}
\authornotemark[1]
\affiliation{
  \institution{Zhejiang University}
  \city{Hangzhou}
  \country{China}
}
\email{yuanchenbei@zju.edu.cn}

\author{Hao Chen}
\authornotemark[1]
\affiliation{
  \institution{The Hong Kong Polytechnic University}
  \city{Hong Kong}
  \country{China}
}
\email{sundaychenhao@gmail.com}

\author{Qijie Shen}
\affiliation{
  \institution{Alibaba Group}
  \city{Hangzhou}
  \country{China}
}
\email{qijie.sqj@alibaba-inc.com}

\author{Zheng Yuan}
\affiliation{
  \institution{The Hong Kong Polytechnic University}
  \city{Hong Kong}
  \country{China}
}
\email{yzheng.yuan@connect.polyu.hk}

\author{Huan Gong}
\affiliation{
  \institution{National University of Defense Technology}
  \city{Changsha}
  \country{China}
}
\email{gongh15@outlook.com}

\author{Senzhang Wang}
\affiliation{
  \institution{Central South University}
  \city{Changsha}
  \country{China}
}
\email{szwang@csu.edu.cn}

\author{Feiran Huang}
\authornote{Corresponding author.}
\affiliation{
  \institution{Jinan University}
  \city{Guangzhou}
  \country{China}
}
\email{huangfr@jnu.edu.cn}

\author{Xiao Huang}
\affiliation{
  \institution{The Hong Kong Polytechnic University}
  \city{Hong Kong}
  \country{China}
}
\email{xiaohuang@comp.polyu.edu.hk}







\renewcommand{\shortauthors}{Yijie Zhang, et al.}

\begin{abstract}

Representing information of multiple behaviors in the single graph collaborative filtering (CF) vector has been a long-standing challenge. This is because different behaviors naturally form separate behavior graphs and learn separate CF embeddings. Existing models merge the separate embeddings by appointing the CF embeddings for some behaviors as the primary embedding and utilizing other auxiliaries to enhance the primary embedding. However, this approach often results in the joint embedding performing well on the main tasks but poorly on the auxiliary ones. To address the problem arising from the separate behavior graphs, we propose the concept of \textbf{\underline{P}artial \underline{O}rder Recommendation \underline{G}raphs (POG)}. POG defines the partial order relation of multiple behaviors and models behavior combinations as weighted edges to merge separate behavior graphs into a joint POG. Theoretical proof verifies that POG can be generalized to any given set of multiple behaviors. Based on POG, we propose the tailored \textbf{\underline{P}artial \underline{O}rder \underline{G}raph \underline{C}onvolutional \underline{N}etworks (POGCN)} that convolute neighbors' information while considering the behavior relations between users and items. POGCN also introduces a partial-order BPR sampling strategy for efficient and effective multiple-behavior CF training. POGCN has been successfully deployed on the homepage of Alibaba for two months, providing recommendation services for over one billion users. Extensive offline experiments conducted on three public benchmark datasets demonstrate that POGCN outperforms state-of-the-art multi-behavior baselines across all types of behaviors. Furthermore, online A/B tests confirm the superiority of POGCN in billion-scale recommender systems.

\end{abstract}





\maketitle

\section{Introduction}
Graph-based collaborative filtering (CF) models \cite{wu2022grecsurcey, guo2020gksurcey, wang2021glrec} have emerged as the state-of-the-art in predicting user-item interactions, particularly for single behaviors like clicks \cite{he2020lightgcn, chen2024macro, wu2021sgl, yu2022simgcl}. However, in real-world recommender systems, there are other behaviors equally crucial as clicks, such as favors and purchases, which greatly impact user retention and platform revenue~\cite{jin2020mbgcn,xia2021mbgmn}. Nonetheless, CF models focused on a single behavior may not perform well for other behaviors. For example, graph embeddings trained on click data often yield suboptimal results for favor and purchase recommendations, and vice versa. This phenomenon requires modern recommender systems to train respective CF embeddings for every behavior, leading to considerable time, cost, and storage burdens. Hence, it is crucial to design an effective multi-behavior graph-based CF model that can efficiently recommend multiple behaviors using a single CF embedding.

Existing multi-behavior CF models usually merge the separate embeddings by affording certain behaviors higher priority and serving others as secondary to enhance the primary embedding. For instance, NMTR~\cite{gao2019nmtr}, which extends the neural CF model~\cite{he2017ncf}, considers purchasing behavior as the primary behavior and incorporates other behaviors to enhance purchasing predictions. Similarly, MB-CGCN~\cite{cheng2023mbcgcn} builds upon NMTR by integrating LightGCN~\cite{he2020lightgcn} to improve graph embeddings for purchasing by leveraging click and cart addition embeddings. On the other hand, IMGCF~\cite{zhang2023imgcf} adopts a different approach by learning separate embeddings for each behavior graph and then averaging these for multi-behavior recommendations. GHCF~\cite{chen2021ghcf} introduces a unique weighting strategy for combining embeddings from different behavior graphs. Another line of multi-behavior research, Click Through Rate (CTR) prediction models like ESMM~\cite{ma2018essm}, MMOE~\cite{ma2018mmoe}, and PLE~\cite{tang2020ple} relies on designing complex MLP structures to deal with the multi-behavior recommendation but they fail to represent users and items with single embedding vectors. 

\begin{figure}[tbp]
    \centering
    \includegraphics[width=\linewidth, trim=0cm 0cm 0cm 0cm,clip]{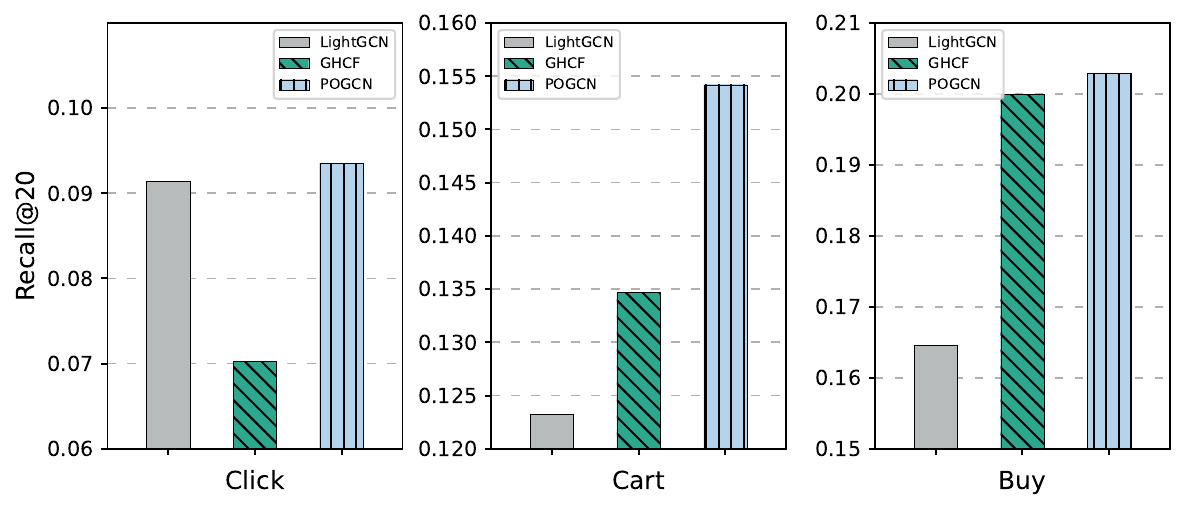}
    \caption{Illustration of the seesaw problems of current multi-behavior collaborative filtering models.}
    \label{fig:intro}
\end{figure}

However, current multi-behavior collaborative filtering (CF) models still learn the embedding of different behaviors separately, raising severe seesaw problems. The merged embedding may perform well on the priority behavior but sacrifice recommendation performance on the auxiliaries. As shown in Figure.\ref{fig:intro}, LightGCN performs well on click recommendations but performs significantly worse on the other two behaviors. On the other hand, GHCF, which treats ``buy'' as the priority, outperforms LightGCN in add-to-cart and purchase recommendations. However, GHCF fails in recommending clicks. In practical applications, all types of behaviors have significant implications for user experience and platform revenue. As stated, predicting clicks affects user intentions such as activation, while add-to-cart and purchasing impact the platform's revenue. Therefore, it is crucial to model the separate multi-behaviors in a joint structure and thus train single embedding naturally.

However, encoding the separate multiple behaviors into a joint structure poses the following three challenges:
\begin{itemize}
    \item \textbf{Separate Graphs}: The current graph concept cannot represent multiple behaviors within one graph. Defining a suitable joint graph to simultaneously organize arbitrary multiple behaviors is challenging.
    \item \textbf{Complicated Behavior Relations}: Multiple behaviors can have complex combinations. For instance, users may directly purchase items without favoriting or adding them to the cart. Alternatively, they may add several items to carts before making a purchase. Thus, modeling the combination of behaviors is more difficult than modeling individual behaviors.
    \item \textbf{Behavior Combination Ordering}: Existing models only provide the order of behaviors. It's challenging to determine the order of behavior combinations such as comparing the order of ``buy'', ''click\&favor\&cart'', and ``buy\&click''.
\end{itemize}

In this paper, we tackle the above three challenges by upgrading the infrastructure of multiple behavior graph collaborative filtering and introducing the general \textbf{\underline{P}artial \underline{O}rder Recommendation \underline{G}raph (POG)} to merge separate multi-behavior graphs into a unified one. Specifically, POG utilizes a ``graded partial order set'' to model all the potential behavior combinations between users and items. Consequently, POG is able to convert any behavior combination into a joint weighted graph. With the help of POG, we propose tailored \textbf{\underline{P}artial \underline{O}rder \underline{G}raph \underline{C}onvolutional \underline{N}etworks (POGCN)} to train a single embedding that can benefit multi-behavior recommendation simultaneously. Additionally, we simplify the original multiple-behavior BPR sampling process by developing a tailored partial order BPR sampling strategy. As depicted in Figure.\ref{fig:intro}, POGCN naturally resolves the seesaw problem and improves the recommendation performance on ``click'', ``cart'', and ``buy'' simultaneously. Our paper's primary contributions can be summarized as follows:
\begin{itemize}
    \item We formally define the partial order recommendation graph to describe the complicated behavior combinations for the seesaw phenomenon in multi-behavior recommendation naturally. Moreover, we prove the completeness of the definition of the POG that can deal with arbitrary multiple behaviors.   
    \item We propose a novel partial order graph convolutional network to learn representative single embedding for multi-behavior CF tasks. Besides, we propose a simplified partial order BPR sampling strategy\footnote{Source codes are available at \url{https://github.com/Wings236/POGCN}.}.
    \item POGCN has been serving as a core recall model at the homepage of one of the biggest e-commerce platforms---Alibaba, offering accurate and suitable recommendations to over one billion users.
    \item Extensive experiments on three benchmark datasets present that \model outperforms the state-of-the-art multi-behavior collaborative filtering methods for above 16.84\% Recall, and 19.67\% NDCG on average. Online A/B test also demonstrates that \model brings 2.02\% UCTR and 2.84\% GMV improvement in industrial platforms.
\end{itemize}

\section{PRELIMINARY}
In this section, we begin by defining the problem of multi-behavior collaborative filtering. We then proceed to introduce the definition of the partial order of behaviors and subsequently define the partial order for combinations of behaviors. Finally, we provide the definition of the graph that will be used in subsequent sections, which includes the ``separate behavior graph'', the ``behavior combination graph'', and our ``partial order recommendation graph (POG)''.

\subsection{Notations and Problem Definition}
\paragraph{\textbf{Notations}} We use $\mathcal{U}$, $\mathcal{I}$, $\mathcal{B}$, and $\gC$ to represent the sets of users, items, behaviors, and behavior combination set, respectively. Here, $M$ represents the number of users, $N$ represents the number of items, $K$ represents the number of behavior categories, and $H$ represents the number of behavior combinations. Given $\gR$ as the set of all interactions of all behaviors, $\gR_k$ denotes the interactions of the $k$-th behavior. Then, describing from the behavior aspect, for each behavior, we have a matrix $\mR_k$ to describe the interaction of the $k$-th behavior. Specifically, $\mR_{k_{ui}}=1$ if user $u$ and item $i$ have the $k$-th behavior, and $\mR_{k_{ui}}=0$ if user $u$ and item $i$ do not have the $k$-th behavior. Describing from the users and item interaction aspect, $\mB_{ui} = \{b_{ui,1}, \dots, b_{ui,K_{ui}}\}$ denotes the set of behaviors between user $u$ and item $i$, and $K_{ui}$ denotes the number of behaviors between user $u$ and item $i$, where $b_{ui,1}, \dots, b_{ui,K_{ui}} \in \mathcal{B}$, and $\mB_{ui}$ is an element of the behavior combination set $\gC$.

\paragraph{\textbf{Multi-Behavior Collaborative Filtering}} The ultimate objective of the multi-behavior~(embedding-based) collaborative filtering~\cite{he2017ncf,wang2019ngcf,he2020lightgcn} model is to acquire a single representation embedding for each user and item, denoted as $\mE_\gU$ and $\mE_{\gI}$. On a micro level, we utilize $\ve_u$ and $\ve_i$ to represent the trained embedding vectors for user $u\in\gU$ and item $i\in\gI$. 

During the inference step, the relationship between a given user and item pair can be calculated by multiplying their embeddings, $\hat{Y}_{ui} = \ve_u^\top \cdot \ve_i$. 
Subsequently, for any given user, the online recommender system will rank all the items to filter the most relevant ones. The filtered items for user $u$ can be formally defined as,
\begin{equation}
\label{eq:cf}
    \hat{\gY}_{u} = Filter(\{\hat{Y}_{ui}, \  \forall \ i \in \gI, N_{\text{top}}\}),
\end{equation}  
where $N_{\text{top}}$ represents the hyperparameter that controls the number of filtered items.
During the evaluation process, the recommender system assesses the recommendation performance of all behaviors based on $\hat{\gY}_{u}$.

\subsection{Definition of Behavior Relations}

\paragraph{\textbf{Partial order of Behaviors}} The purpose of proposing a partial order is to address situations where the recommender system cannot determine the order of two or more behaviors. For example, consider the comparison of ``favorite'', ``share'', and ``adding cart''. Previously, the order of behavior definitions would typically treat ``favorite'', ``share'', and ``adding cart'' as the same behavior, such as ``indirect behavior'', without being able to distinguish between different behaviors. On the other hand, our proposed partial order of behaviors can accommodate ambiguous orders.

\begin{definition}[\textit{Partial Order of Behaviors}] \label{def1}
We define a binary relation \( \leq_b \) on behavior set \( \gB \), such that for all behaviors \( x, y, z \in \gB \), the following conditions are satisfied:
\begin{itemize}
    \item \textit{Reflexivity:} For every \( x \in \gB \), \( x \leq_b x \).
    \item \textit{Anti-symmetry:} For all \( x, y \in \gB \), if \( x \leq_b y \) and \( y \leq_b x \), then \( x = y \).
    \item \textit{Transitivity:} For all \( x, y, z \in \gB \), if \( x \leq_b y \) and \( y \leq_b z \), then \( x \leq_b z \).
\end{itemize}
\end{definition}

After the definition of the partial order of behaviors, the above example can be expressed as the following relation: \[click \le_b share, favorite, adding\ cart \le_b buy,\]
where ``$share, favorite, adding\ cart$'' can be called the incomparable relationship in the definition of partial order.

\paragraph{\textbf{Rank Functions of Behaviors}}
By defining the partial order of behaviors, we can establish the partial order of behavior combinations based on the nature of the partial order set. To facilitate subsequent numerical computations, we also need to introduce the concept of graded partial order of behaviors.

\begin{definition}[\textit{Graded Partial Order of Behavior}] \label{def2}
The graded partial order of behavior is defined as a partial order set \((\gB, \leq_b)\), augmented with a rank function \(\rho_b: \gB \to \mathbb{N}^+\), where \(\mathbb{N}^+\) represents the set of positive natural numbers. The function $\rho_b$ satisfies the following conditions:
\begin{itemize}
    \item \textit{Order-Preserving}: For all \( x, y \in \gB \), if \( x \leq_b y \), then \(\rho_b(x) \leq \rho_b(y)\).
    \item \textit{Covering Condition}: For all \( x, y \in \gB \), if \( y \) covers \( x \) (i.e., there is no \( z \in \gB\) such that \( x \leq_b z \leq_b y,\) and  \(z\ne x, z\ne y  \)), then \(\rho_b(y) = \rho_b(x) + 1\).
\end{itemize}
\end{definition}

With the definition of a graded partial order of behavior, any arbitrary partial order of behaviors can be represented by a corresponding positive integer. 

\begin{figure*}[t]\label{framework}
\centering
    \includegraphics[width=\linewidth, trim=0 180 0 170, clip]{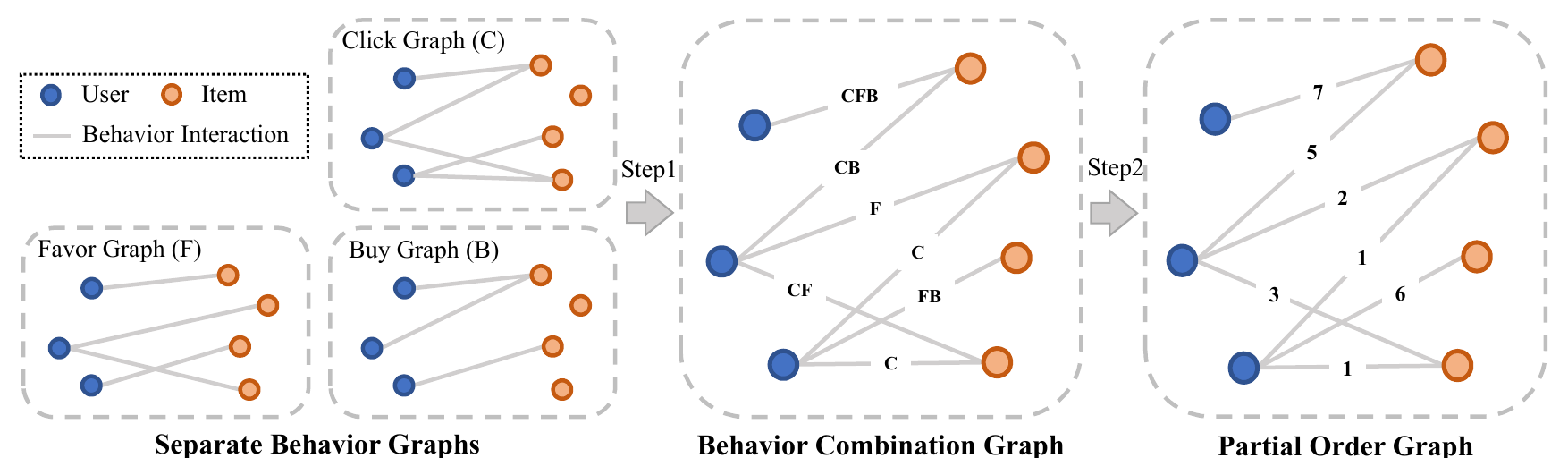}
    \caption{An illustration of the construction of Partial Order Recommendation Graph~(POG). The partial order of behaviors is given as \textit{``Click(C) $\le_b$ Favor(F) $\le_b$ Buy(B)''}. According to Definition~\ref{def3}, the rank function of the partial order recommendation graph can be given as $\rho_c(\{\textrm{C, F, B}\})=7$, $\rho_c(\{\text{F, B}\})=6$, $\rho_c(\{\text{C, B}\})=5$, $\rho_c(\{\text{B}\})=4$, $\rho_c(\{\text{C, F}\})=3$, $\rho_c(\{\text{F}\})=2$, $\rho_c(\{\text{C}\})=1$.}
\label{fig:framework}
\end{figure*}


In~\autoref{sec:comp_combination proof}, we provide the proof of completeness for the grade partial order of behavior combinations to showcase the generality of our definition.

\subsection{Definition of Graphs}
In this subsection, we formally define the graphs that will be used in the proceeding contents, which are also presented in~\autoref{fig:framework}. We begin by discussing the separate behavior graph utilized in related works. Then, we propose the behavior combination graph to depict all combinations of behaviors in a single graph. Finally, we present the partial order graph to assign a rank value to each combination.

\paragraph{\textbf{Separate Behavior Graphs.}} Separate behavior graphs utilize $K$ different graphs to describe each users' behavior, namely $\gG_{sep} = \{\gG_1,\cdots,\gG_K\}$. 
For different behaviors $k$, we have corresponding historical interaction information $\gR_k$ between users and items. Therefore, we can represent the corresponding behavior graph of behavior $k$ as $\gG_k=(\gU,\gI,\gR_k)$. Besides, we utilize $\mR_k$ to denote the interaction matrix of $k$-th behavior between users and items.

\paragraph{\textbf{Behavior Combination Graph.}}
The behavior combination graph's edges describe the combined behavior situations $\mB_{ui}$, representing the set of behaviors occurring between user $u$ and item $i$. For instance, if user $u$ clicks and buys item $i$, then the edge between  user $u$ and item $i$ is $\mB_{ui}=\{click, buy\}$. In this context, we define the behavior combination graph as $\gG_{com} = (\gU, \gI, \gR)$.

\paragraph{\textbf{Partial Order Recommendation Graph.}} 
Since the edges in the behavior combination graph represent discrete set values that cannot be ordered as continuous values, we utilize the rank function in Definition~\autoref{def2} to convert these discrete sets into ordered sets. We refer to this graph as a behavior combination partial order graph or simply a \textbf{Partial Order Recommendation Graph (POG)}. Formally, it is defined as $\gG_{\text{po}} = (\gU,\gI,\rho_c(\gR))$, where $\rho_c(\cdot)$ denotes the rank function to convert the behavior combinations to integers. $\mR$ denotes the interaction matrix of the partial order graph between users and items.

\section{METHODOLOGY}

In this section, we begin by explaining the process of generating the Partial Order Recommendation Graph~(POG) from the original separate behavior graphs. Subsequently, we delve into the graph convolution on the POG. Lastly, we provide the formula for the partial order BPR loss for multi-behavior recommendation.

\subsection{Construction of POG}
The construction of POG involves one predefinition and one converting step. In the predefinition step, the partial order of behaviors and the partial order of behavior combinations are defined. As depicted in~\autoref{fig:framework}, once the orders are defined, separate behavior graphs can be transformed into a behavior combination graph, and subsequently into the partial order recommendation graph.

\subsubsection{Predefinition} As defined in Definition~\autoref{def1}, exports from recommender systems can first define the partial order of behaviors in a customized mode. As shown in~\autoref{fig:framework}, considering three behaviors $\gB = \{click, favor, buy\}$, their partial order $\le_b $ can be defined as ``$click \le_b favor \le_b buy $''.

Then, with Definition \ref{def2}, we can define the rank function $\rho_c$ to map any behavior combination $\gC_i$~(subset of the behavior set $\gB$) to an integer. Specifically, the rank of each set can be defined using the following binary comparison strategy $\le_c$:

\begin{definition}[\textit{Behavior Combination Binary Relation}] \label{def3}
We define a new binary relation \( \leq_c \) between \( \gC_i \) and \(\gC_j \). \( \gC_i \) and  \( \gC_j \) represent the behavior combination.
For convenience, we define $f(k,\mathcal{C}_i)$ as the function that returns the number of $k$-ranked behavior ($\rho_b(b_s)=k$) in $\mathcal{C}_i$. $f(k,\mathcal{C}_i)$ can be formally defined as $f(k,\mathcal{C}_i) = |\{b_s \mid \rho_b(b_s) = k, b_s \in \mathcal{C}_i\}|$, where $|\cdot|$ denotes the number count function.
The relation between behavior combinations can be computed in the following recursion formula:




\begin{enumerate}
    \item \textbf{Equality Check:} If \( \gC_i = \gC_j \), then \(\gC_i \leq_c \gC_j \); otherwise, let \( k\) be the max value of $\rho_b$, and proceed to the next step.
    \item \textbf{Behavior Intensity Count Comparison:} If \( f(k,\gC_i) < f(k,\gC_j) \) (or \( f(k,\gC_j) < f(k,\gC_i) \)), then \( \gC_i \leq_c \gC_j \) (or \( \gC_j \leq_c \gC_i \) ); otherwise, proceed to the next step.
    \item \textbf{Decrease Intensity:} Decrease \( k \) by 1 (i.e., \( k = k - 1 \)) and repeat from step 2 until \( k = 1 \).
    \item \textbf{Incomparability Determination:} If \( k = 1 \) and \( f(k,\gC_i) = f(k,\gC_j) \), then \( \gC_i \) and \( \gC_j \) are considered incomparable.
\end{enumerate}
\end{definition}

With the above definition of ``Behavior Combination Binary Relation'', we have the following corollary to demonstrate that the behavior combination set $\gC$ can also have a rank function to map any given behavior combination to an integer:
\begin{corollary} \label{cor1}
The set \((\gC, \leq_c)\), when equipped with a rank function \(\rho_c: \gC \to \mathbb{N}^+\), constitutes a graded partial order set.
\end{corollary}

A detailed proof of Corollary~\autoref{cor1}, under complex situations, can be found in Appendix \ref{sec:comp_combination proof}.

\paragraph{\textbf{Rank function example.}} \autoref{fig:framework} presents a specific example of a rank function. For convenience, we use the abbreviations ``B'', ``F'', and ``C'' to represent ``buy'', ``favor'', and ``click''  respectively in the presentation of the rank function. Given the partial order set and Definition~\autoref{def3}, the rank function of the partial order recommendation graph can be defined as follows: $\rho_c(\{\text{C, F, B}\})=7$, $\rho_c(\{\text{F, B}\})=6$, $\rho_c(\{\text{C, B}\})=5$, $\rho_c(\{\text{B}\})=4$, $\rho_c(\{\text{C, F}\})=3$, $\rho_c(\{\text{F}\})=2$, $\rho_c(\{\text{C}\})=1$. By examining the rank function, we can observe that ``buy'' is the most important behavior, and therefore, combinations that include ``buy'' have a higher rank than those without it. Furthermore, based on the comparison of the most important behavior ``buy'', $\{\text{B, F}\}$ has a higher rank than $\{\text{B, C}\}$ because ``favor'' is considered more important than ``click''.

\subsubsection{POG Converting}
\autoref{fig:framework} illustrates the entire process of converting the graph from separate behavior graphs to the behavior combination graph and then to the partial order recommendation graph. In summary, the behavior combination graph combines all the edges from the separate behavior graphs, where each edge represents a set of behaviors between a specific user and item pair. Afterward, the partial order recommendation graph assigns an integer weight to the set of behaviors, resulting in a unified weighted graph.
Specifically, given the definition of the behavior combination between user $u$ and item $i$ as $\mB_{ui}$ and the rank function $\rho_c(\cdot)$, then the interaction matrix of the partial order recommendation graph can be defined as:

\begin{equation}
\label{eq:pog}
\mR_{ui} = 
\begin{cases} 
\rho_c(\mB_{ui})^{\tau} &,   \text{if } \mB_{ui} \neq \phi; \\
0 &,  \text{if } \mB_{ui} = \phi,\\
\end{cases}
\end{equation}
where $\tau>0$ represents the temperature coefficient~\cite{van1987SimulatA,bertsimas1993SimulatA2} used to adjust the importance weight of various behavior combinations. When $\tau$ is increased, the distance between behavior combinations also increases, while decreasing $\tau$ results in a smaller distance between combinations. If $\tau = 0$, the partial order interaction matrix assigns a value of 1 to all interacting user-item pairs, indicating that all behavior combinations are considered equally important.

\subsection{Partial Order Graph Convolutional Networks}
In contrast to GHCF~\cite{chen2021ghcf}, MB-CGCN~\cite{cheng2023mbcgcn}, and IMGCF~\cite{zhang2023imgcf}, POGCN train a single graph-based CF embedding on the POG. In this section, we first present POGCN in matrix format and then provide the message passing formula to enhance its applicability in the industry.


 For any given user $u$ and item $i$, we first initialized the original embedding vectors $\ve_u^{(0)}, \ve_i^{(0)} \in \mathbb{R}^d$, where $d$ represents the dimension of the vectors. Then we have an original embedding matrix for all users and items: $\mE^{(0)}=[\ve^{(0)}_{u_1}, \dots, \ve^{(0)}_{u_M}, \ve^{(0)}_{i_1}, \dots, \ve^{(0)}_{i_N}]^T$.

\paragraph{\textbf{Matrix Form}} Given the interaction matrix of partial order recommendation graph $\mR$, we extend the interaction matrix to define the partial order adjacent matrix:
\begin{equation}
\mA = 
\begin{bmatrix}
0 & \mR \\
\mR^T & 0
\end{bmatrix},
\end{equation}
then we are able to obtain the partial order graph convolution formula as:
\begin{equation}
\mE^{(l+1)} = (\mD^{-\frac{1}{2}}\mA\mD^{-\frac{1}{2}})\mE^{(l)},
\end{equation}
where $\mD \in \mathbb{R}^{(M+N) \times (M+N)}$ is a degree matrix, and $\mD_{ii}=\sum_{j}\mA_{ij}$, which denotes the sum of $i$-th row value of the partial order adjacent matrix $\mA$. 
With the above definition,  the final partial order embedding matrix can be computed as follows:
\begin{equation}
    \mE = \frac{1}{L+1}\sum_{l=0}^{L} \mE^{(l)}.
\end{equation}
After the graph convolution, POGCN will naturally get a single CF embedding to achieve the CF task defined in EQ. (\ref{eq:cf}).

\paragraph{\textbf{Message Passing Form}} The partial order recommendation graph can also be done in a neighbor-wise message passing form. Specifically, the $l$-th message passing of user $u$ and item $i$ can be given as:
\begin{equation}
\begin{aligned}
\ve_{u}^{(l+1)} &= \sum_{i \in \mathcal{N}_u} \frac{\mR_{ui}}{\sqrt{\sum_t \mR_{ut}} \sqrt{\sum_t \mR_{ti}}}\ve_{i}^{(l)}, \\
\ve_{i}^{(l+1)} &= \sum_{u \in \mathcal{N}_i} \frac{\mR_{ui}}{\sqrt{\sum_t \mR_{ti}} \sqrt{\sum_t \mR_{ut}}}\ve_{u}^{(l)},
\end{aligned}
\end{equation}
where $\mathcal{N}_{u}$ and $\mathcal{N}_{i}$ represent the sets of neighboring nodes for user $u$ and item $i$, respectively. $\sum_t \mR_{ut}$ and $\sum_t \mR_{ti}$ are the sums of the $u$-th row and the $i$-th column in the partial order interaction matrix $\mR$, respectively. $\frac{\mR_{ui}}{\sqrt{\sum_t \mR_{ut}} \sqrt{\sum_t \mR_{ti}}}$ is the Laplace normalization of message passing in order to avoid numerical instabilities and exploding/vanishing gradients~\cite{kipf2016semigcn}. 

After performing $L$ rounds of propagation, we take the average of the obtained partial order embeddings at each layer to obtain the final representation:
\begin{equation}
\ve_u = \frac{1}{L + 1} \sum\limits_{l=0}^L \ve_u^{(l)},
\ve_i = \frac{1}{L + 1} \sum\limits_{l=0}^L \ve_i^{(l)},
\end{equation}
where $\ve_u$ and $\ve_i$ represent the final POGCN embedding of user $u$ and item $i$ respectively.

\subsection{Partial Order Training Strategy}
Bayesian Personalized Ranking (BPR) has been extensively applied in collaborative filtering. Traditional BPR mainly concentrates on training with a single behavior. In this subsection, we introduce an extension called partial order BPR, which aims to achieve efficient and effective multi-behavior collaborative filtering.

\paragraph{\textbf{Traditional Multi-behavior BPR}} Intuitively, one traditional way to combine the BPR loss for multiple behaviors is to apply BPR separately for each behavior and then assign different weights to each behavior. This can be represented as follows:
\begin{equation}
    \gL_{MTL-BPR} = \sum_{k =1}^K \sum_{\mR_{k_{ui}}=1} \sum_{\mR_{k_{uj}}=0} \alpha_k \cdot \ln (\sigma(\hat{Y}_{ui} - \hat{Y}_{uj})),
\end{equation}
where $\mR_{k_{ui}}=1$ indicates that user $u$ and item $i$ have the $k$-th behavior, while $\mR_{k_{uj}}=0$ indicates that user $u$ and item $j$ do not have the $k$-th behavior. $\alpha_k$ denotes the weight of the $k$-th behavior.

However, the traditional multi-behavior BPR solution has the following limitations:
\textit{1. High training cost}: Performing BPR training requires computing the embedding for one user and two items, and then performing backpropagation. Therefore, performing traditional multi-behavior BPR will incorporate $K$ times of computational cost of single-behavior BPR, thus increasing the computational complexity.
\textit{2. Manual weighting}: In traditional solutions, experts have to manually assign the weight, which cannot automatically adapt according to the importance of behavior combinations.

\paragraph{\textbf{Our POBPR}}To overcome these challenges, we propose leveraging a Multinomial Distribution~\cite{murphy2012machine} $P(\gC) = P(p_1, \cdots, p_H)$ for all $H$ possible behavior combinations, allowing for the sampling of combinations based on their relevance and frequency. Specifically, for any given behavior combination $\gC_h$, the sampling probability is determined as follows:
\begin{equation}
p_h = P(\gC_h) = \frac{\rho_c(\gC_h)^\gamma \cdot num(\gC_h)}{\sum_{j=1}^H \rho_c(\gC_j)^\gamma \cdot num(\gC_j)},
\end{equation}
where $num(\gC_j)$ counts the occurrences of each behavior combination $\gC_j$, and $\gamma$ is a temperature coefficient~\cite{van1987SimulatA, bertsimas1993SimulatA2} that moderates the influence of the rank value of each behavior combination.



\begin{table*}[htbp]
  \centering
  \caption{Multi-behavior recommendation performance comparison results. The best and second-best results in each column are highlighted in \textbf{bold} font and \underline{underlined}, respectively.}
  \resizebox{\linewidth}{!}{
    \begin{tabular}{ccc|cccccc|ccc|cc}
    \toprule
    \multicolumn{3}{c|}{Models} & MC-BPR & NMTR  & GHCF  & MB-GMN & MB-CGCN & IMGCF & ESMM & MMoE & PLE & \model & \textit{Improv.} (\%) \\
    \midrule
    \multicolumn{1}{c}{\multirow{8}[8]{*}{\rotatebox{90}{Beibei}}} & \multicolumn{1}{c}{\multirow{2}[2]{*}{Click}} & Recall & 0.0693 & 0.0325 & 0.0703 & 0.0449 & 0.0495 & 0.0688 & \underline{0.0750}& 0.0685 
& 0.0614 
& \textbf{0.0935} 
& 24.68\%
\\
          &       & NDCG  &       0.0560 &       0.0262 &       0.0617 &       0.0360 &       0.0416 &       0.0571 &       \underline{0.0652}&       0.0588 
&       0.0515 
&       \textbf{0.0841} &  28.95\%
\\
\cmidrule{2-14}          & \multicolumn{1}{c}{\multirow{2}[2]{*}{Cart}} & Recall &       0.1140 &       0.0702 &       0.1347 &       0.0833 &       0.0945 &       0.1338 &       \underline{0.1358}&       0.1240 &       0.1170 
&       \textbf{0.1542} &  13.61\%
\\
          &       & NDCG  &       0.0624 &       0.0397 &       \underline{0.0812} &       0.0443 &       0.0542 &       0.0758 &       0.0806 
&       0.0719 &       0.0672 
&       \textbf{0.0948} &  16.78\%
\\
\cmidrule{2-14}          & \multicolumn{1}{c}{\multirow{2}[2]{*}{Buy}} & Recall &       0.1428 &       0.1222 &       \underline{0.1999} &       0.1220 &       0.1428 &       0.1950 &       0.1680 
&       0.1547 &       0.1789 
&       \textbf{0.2029} &  1.47\%
\\
          &       & NDCG  &       0.0690 &       0.0619 &       \underline{0.1052} &       0.0581 &       0.0724 &       0.0971 &       0.0866 
&       0.0777 &       0.0903 
&       \textbf{0.1083} &  3.01\%
\\
\cmidrule{2-14}          & \multicolumn{1}{c}{\multirow{2}[2]{*}{Mean}} & Recall &       0.1087 &       0.0750 &       \underline{0.1350} &       0.0834 &       0.0956 &       0.1326 &       0.1262 
&       0.1157 &       0.1191 
&       \textbf{0.1502} &  11.28\%
\\
          &       & NDCG  &       0.0624 &       0.0426 &       \underline{0.0827} &       0.0461 &       0.0561 &       0.0767 &       0.0775 
&       0.0694 &       0.0696 
&       \textbf{0.0957} &  15.80\%
\\
    \midrule
    \multicolumn{1}{c}{\multirow{10}[10]{*}{\rotatebox{90}{Taobao}}} & \multicolumn{1}{c}{\multirow{2}[2]{*}{Click}} & Recall &       0.0180 &       0.0016 &       \underline{0.0392} &       0.0006 &       0.0018 &       0.0177 &       0.0070 
&       0.0014 &       0.0013 
&       \textbf{0.0424} &  8.34\%
\\
          &       & NDCG  &       0.0114 &       0.0009 &       \underline{0.0274} &       0.0003 &       0.0012 &       0.0113 &       0.0043 
&       0.0008 &       0.0008 
&       \textbf{0.0281} &  2.46\%
\\
\cmidrule{2-14}          & \multicolumn{1}{c}{\multirow{2}[2]{*}{Cart}} & Recall &       0.0276 &       0.0019 &       \underline{0.0498} &       0.0012 &       0.0031 &       0.0327 &       0.0141 
&       0.0021 &       0.0016 
&       \textbf{0.0555} &  11.39\%
\\
          &       & NDCG  &       0.0116 &       0.0006 &       \underline{0.0225} &       0.0004 &       0.0016 &       0.0136 &       0.0059 
&       0.0007 &       0.0009 
&       \textbf{0.0244} &  8.40\%
\\
\cmidrule{2-14}          & \multicolumn{1}{c}{\multirow{2}[2]{*}{Favor}} & Recall &       0.0271 &       0.0015 &       \underline{0.0595} &       0.0005 &       0.0013 &       0.0277 &       0.0054 
&       0.0026 &       0.0011 
&       \textbf{0.0633} &  6.52\%
\\
          &       & NDCG  &       0.0116 &       0.0004 &       \underline{0.0252} &       0.0003 &       0.0004 &       0.0122 &       0.0021 
&       0.0009 &       0.0003 
&       \textbf{0.0261} &  3.83\%
\\
\cmidrule{2-14}          & \multicolumn{1}{c}{\multirow{2}[2]{*}{Buy}} & Recall &       0.0172 &       0.0027 &       \underline{0.0324} &       0.0021 &       0.0008 &       0.0309 &       0.0072 
&       0.0005 &       0.0027 
&       \textbf{0.0396} &  22.07\%
\\
          &       & NDCG  &       0.0067 &       0.0008 &       \underline{0.0155} &       0.0007 &       0.0002 &       0.0128 
&       0.0027 
&       0.0002 &       0.0015 
&       \textbf{0.0188 
}&  21.04\%\\
\cmidrule{2-14}          & \multicolumn{1}{c}{\multirow{2}[2]{*}{Mean}} & Recall &       0.0225 &       0.0019 &       \underline{0.0452} &       0.0011 &       0.0017 &       0.0272 &       0.0084 
&       0.0017 &       0.0017 
&       \textbf{0.0502} &  11.05\%
\\
          &       & NDCG  &       0.0103 &       0.0007 &       \underline{0.0226} &       0.0004 &       0.0009 &       0.0125 &       0.0038 
&       0.0006 &       0.0009 
&      \textbf{0.0243}&  7.49\%\\
    \midrule
    \multirow{10}[10]{*}{\rotatebox{90}{Tenrec}} & \multirow{2}[2]{*}{Click} & Recall &       0.1284 &       0.0066 &       \underline{0.1410} &       0.0102 &       0.0744 &       0.0449 &       0.0108 
&       0.0224 &       0.0121 
&       \textbf{0.1627} &  15.36\%
\\
          &       & NDCG  &       0.0827 &       0.0038 &       \underline{0.0943} &       0.0055 &       0.0511 &       0.0290 &       0.0069 
&       0.0137 &       0.0075 
&       \textbf{0.1097} &  16.35\%
\\
\cmidrule{2-14}          & \multirow{2}[2]{*}{Like} & Recall &       \underline{0.1048} &       0.0105 &       0.0941 &       0.0128 &       0.0756 &       0.0661 &       0.0173 
&       0.0222 &       0.0158 
&       \textbf{0.1221} &  16.44\%
\\
          &       & NDCG  &       \underline{0.0488} &       0.0038 &       0.0393 &       0.0041 &       0.0316 &       0.0325 &       0.0063 
&       0.0091 &       0.0076 
&       \textbf{0.0621} &  27.33\%
\\
\cmidrule{2-14}          & \multirow{2}[2]{*}{Share} & Recall &       \underline{0.1176} &       0.0147 &       0.1029 &       0.0294 &       0.0882 &       0.0772 &       0.0294 
&       0.0147 &       0.0404 
&       \textbf{0.1324} &  12.50\%
\\
          &       & NDCG  &       0.0399 &       0.0035 &       \underline{0.0453} &       0.0091 &       0.0346 &       0.0400 &       0.0095 
&       0.0036 &       0.0145 
&       \textbf{0.0636} &  40.25\%
\\
\cmidrule{2-14}          & \multirow{2}[2]{*}{Follow} & Recall &       0.0289 &       0.0051 &       \underline{0.0408} &       0.0090 &       0.0405 &       \underline{0.0408} &       0.0357 
&       0.0204 &       0.0102 
&       \textbf{0.0697} &  70.83\%
\\
          &       & NDCG  &       \underline{0.0279} &       0.0015 &       0.0225 &       0.0021 &       0.0202 &       0.0162 &       0.0129 
&       0.0128 &       0.0034 
&       \textbf{0.0380} &  36.25\%
\\
\cmidrule{2-14}          & \multirow{2}[2]{*}{Mean} & Recall &       \underline{0.0949} &       0.0092 &       0.0947 &       0.0154 &       0.0697 &       0.0573 &       0.0233 
&       0.0199 &       0.0196 
&       \textbf{0.1217} &  28.19\%
\\
          &       & NDCG  &       0.0498 &       0.0031 &       \underline{0.0503} &       0.0052 &       0.0344 &       0.0294 &       0.0089 
&       0.0098 &       0.0083 
&       \textbf{0.0683} &  35.74\%
\\
    \bottomrule
    \end{tabular}%
    }
  \label{tab:main_result}%
\end{table*}%

With the above definition of the categorical distribution, we can provide the complete formula for the POBPR loss as follows:
\begin{equation}
\mathcal{L}_{POBPR} = \mathbb{E}_{\gC_h \sim P(\gC)} \left[ \sum_{(u,i) \in \mathcal{D}_h^+} \sum_{j \in \mathcal{D}_{u}^-} \ln \sigma(\hat{Y}_{ui} - \hat{Y}_{uj}) \right],
\end{equation}
where $\mathbb{E}_{\gC_h \sim P(\gC)}$ represents the expectation over the distribution of behavior combinations, $\mathcal{D}_h^+$ denotes the sets of positive user-item pairs under the behavior combination $\gC_h$, and $\mathcal{D}_{u}^-$ denotes the set of negative items under all interactions of user $u$.
We further demonstrate the equivalence between the POBPR and the original separate formula of multi-behavior in terms of the maximum likelihood in Appendix \ref{sec:po-BPR}.

\section{EXPERIMENTS}

In this section, we conduct both offline and online experiments, aiming to answer the following research questions.
\begin{itemize}
    \item \textbf{RQ1:} Does \model outperform current state-of-the-art recommendation models?
    \item  \textbf{RQ2:} What are the effects of different components in \model?
    \item \textbf{RQ3:} How do key hyper-parameters impact \model?
    \item \textbf{RQ4:} How does \model perform on real-world industrial recommender systems?
\end{itemize}

\subsection{Experimental Setup}
\subsubsection{Datasets}
We conduct comprehensive experiments on three widely used benchmark datasets \textbf{Beibei}~\cite{gao2019nmtr}, \textbf{Taobao}~\cite{zhu2018taobao}, and \textbf{Tenrec}~\cite{yuan2022tenrec} including both e-commerce and content recommendation scenarios for offline evaluation to verify the effectiveness and universality of \model. 
The statistics of these datasets are shown in Table \ref{tab:stats}.
In order to avoid the cold start situation of interaction, following previous works~\cite{wang2019ngcf,he2020lightgcn}, we filter out at least 10 interactive items and users to conduct the experiments. 
Dataset statistics are demonstrated in~\autoref{tab:stats}.
We further describe the details of these datasets in Appendix \ref{sec:data_appendix}.
\begin{table}[htbp]
  \centering
  \caption{Statistics of the experimental datasets.}
  \resizebox{\linewidth}{!}{
    \begin{tabular}{c|c|c|c|c|c|c}
    \toprule
    Datasets & Users & Items & Click & Cart/Like & Favor/Share & Buy/Follow \\
    \midrule
    Beibei & 21,716 & 7,977 & 2,412,586 & 642,622 & \textbackslash{} & 304,576 \\
    Taobao & 26,213 & 64,822 & 1,341,843 & 52,289 & 23,676 & 20,880 \\
    Tenrec & 19,035 & 15,539 & 1,367,967 & 11,295 & 1,503 & 1,330 \\
    \bottomrule
    \end{tabular}%
    }
  \label{tab:stats}%
\end{table}%

\subsubsection{Compared Baselines}
To comprehensively verify the effectiveness of \model. We compare our proposed \model with six multi-behavior CF models, three multi-task recommendation models, and four single-behavior CF models. (I) \textbf{Multi-behavior CF models}: \textit{MC-BPR}~\cite{loni2016mcbpr}, \textit{NMTR}~\cite{gao2019nmtr}, \textit{GHCF}~\cite{chen2021ghcf}, \textit{MB-GMN}~\cite{xia2021mbgmn}, \textit{MB-CGCN}~\cite{cheng2023mbcgcn}, and \textit{IMGCF}~\cite{zhang2023imgcf}.
(II) \textbf{Multi-task recommendation models}: 
\textit{ESMM}~\cite{ma2018essm}, \textit{MMoE}~\cite{ma2018mmoe}, and \textit{PLE}~\cite{tang2020ple}.
(III) \textbf{Single-behavior CF models}: 
\textit{MF-BPR}~\cite{rendle2009bpr}, \textit{NCF}~\cite{he2017ncf}, \textit{NGCF}~\cite{wang2019ngcf}, and \textit{LightGCN}~\cite{he2020lightgcn}.
The details of these baseline models are left in Appendix \ref{sec:baseline_appendix}.

\subsubsection{Implementation Details}
For all models, the embedding size is fixed to 64 and the embedding parameters are initialized with the normal distribution.
The learning rate of \model is searched from \{$1 \times 10^{-3},\ 5 \times 10^{-4},\ 1 \times 10^{-4}$\}, the regularization term is searched from \{$1 \times 10^{-4},\ 5 \times 10^{-5},\ 1 \times 10^{-5}$\}, $\tau$ is searched from [0.2, 1.0] in Taobao and Tenrec with a step of 0.2, and searched from [1.0, 5.0] in Beibei with a step of 1.0, $\gamma$ is searched from [0.2, 2.0] with a step of 0.2. 
The batch size is set to 1024 for all models and the Adam optimizer~\cite{kingma2014adam} is used. For multi-behavior CF models, we adopt the partial order relation ``\textit{click $\le$ cart $\le$ buy}'', ``\textit{click $\le$ cart, favor $\le$ buy}'', and ``\textit{click $\le$ like $\le$ share, follow}'' for Beibei, Taobao, and Tenrec.
For multi-task recommendation models, due to their specific designs for prediction tasks, we replace their backbone with LightGCN to enhance their performance in CF tasks.

\subsubsection{Evaluation Metrics}
Our evaluation adopts a full-ranking evaluation approach, following the state-of-the-art studies~\cite{wang2019ngcf, he2020lightgcn, krichene2020sampledway}.
To evaluate the effectiveness of top-ranked articles, we employ Recall@20 and NDCG@20 as our primary metrics for each type of behavior. In order to facilitate the overall comparison across all types of behaviors, we further adopt the mean metric performance of all behaviors for evaluation. Note that we ran all the experiments five times with different random seeds and reported the average results to prevent extreme cases.

\begin{table}[tbp]
  \centering
  \caption{Comparison results with the single behavior recommendation models.}
  \resizebox{\linewidth}{!}{
    \begin{tabular}{ccc|cccc|c}
    \toprule
    \multicolumn{3}{c|}{Models} & MF-BPR    & NCF   & NGCF  & LightGCN & \model \\
    \midrule
    \multicolumn{1}{c}{\multirow{8}[8]{*}{\rotatebox{90}{Beibei}}} & \multicolumn{1}{c}{\multirow{2}[2]{*}{Click}} & Recall & 0.0635 & 0.0837 & 0.0916 & \underline{0.0920}
& \textbf{0.0935} 
\\
          &       & NDCG  &       0.0506 &       0.0749 &       0.0819 &       \underline{0.0823} 
&  \textbf{0.0841} 
\\
\cmidrule{2-8}          & \multicolumn{1}{c}{\multirow{2}[2]{*}{Cart}} & Recall &       0.0808 &       0.0881 &       0.1210 &       \underline{0.1233} 
&  \textbf{0.1542}
\\
          &       & NDCG  &       0.0435 &       0.0475 &       0.0702 &       \underline{0.0729} 
&  \textbf{0.0948}
\\
\cmidrule{2-8}          & \multicolumn{1}{c}{\multirow{2}[2]{*}{Buy}} & Recall &       0.0870 &       0.1206 &       0.1534 &       \underline{0.1646} 
&  \textbf{0.2029} 
\\
          &       & NDCG  &       0.0415 &       0.0569 &       0.0775 &       \underline{0.0841} 
&  \textbf{0.1083} 
\\
\cmidrule{2-8}          & \multicolumn{1}{c}{\multirow{2}[2]{*}{Mean}} & Recall &       0.0771 &       0.0975 &       0.1220 &       \underline{0.1266}
&  \textbf{0.1502} 
\\
          &       & NDCG  &       0.0452 &       0.0598 &       0.0766 &       \underline{0.0798} 
&  \textbf{0.0957} 
\\
    \midrule
    \midrule
    \multicolumn{1}{c}{\multirow{10}[10]{*}{\rotatebox{90}{Taobao}}} & \multicolumn{1}{c}{\multirow{2}[2]{*}{Click}} & Recall &       0.0191 &       0.0202 &       0.0341 &       \underline{0.0421} 
&  \textbf{0.0424} 
\\
          &       & NDCG  &       0.0121 &       0.0120 &       0.0219 &       \underline{0.0273} 
&  \textbf{0.0281} 
\\
\cmidrule{2-8}          & \multicolumn{1}{c}{\multirow{2}[2]{*}{Cart}} & Recall &       0.0012 &       0.0022 &       0.0062 &       \underline{0.0069} 
&  \textbf{0.0555} 
\\
          &       & NDCG  &       0.0004 &       0.0007 &       0.0025 &       \underline{0.0033} 
&  \textbf{0.0244} 
\\
\cmidrule{2-8}          & \multicolumn{1}{c}{\multirow{2}[2]{*}{Favor}} & Recall &       0.0005 &       0.0032 &       \underline{0.0069} &       0.0054 
&  \textbf{0.0633} 
\\
          &       & NDCG  &       0.0002 &       0.0015 &       \underline{0.0028} &       0.0021 
&  \textbf{0.0261} 
\\
\cmidrule{2-8}          & \multicolumn{1}{c}{\multirow{2}[2]{*}{Buy}} & Recall &       0.0013 &       \underline{0.0066} &       0.0054 &       0.0059 
&  \textbf{0.0396} 
\\
          &       & NDCG  &       0.0005 &       \underline{0.0030} &       0.0021 &       0.0020 
&  \textbf{0.0188}\\
\cmidrule{2-8}          & \multicolumn{1}{c}{\multirow{2}[2]{*}{Mean}} & Recall &       0.0055 &       0.0081 &       0.0132 &       \underline{0.0151} 
&  \textbf{0.0502} 
\\
          &       & NDCG  &       0.0033 &       0.0043 &       0.0073 &       \underline{0.0087} 
&  \textbf{0.0243}\\
    \midrule
    \midrule
    \multirow{10}[10]{*}{\rotatebox{90}{Tenrec}} & \multirow{2}[2]{*}{Click} & Recall &       0.1296 &       0.1258 &       0.1537 &       \underline{0.1620} 
&  \textbf{0.1627} 
\\
          &       & NDCG  &       0.0834 &       0.0814 &       \underline{0.1030}  &       \textbf{0.1097} 
&  \textbf{0.1097} 
\\
\cmidrule{2-8}          & \multirow{2}[2]{*}{Like} & Recall &       0.0032 &       \underline{0.0541} &       0.0247 &       0.0277 
&  \textbf{0.1221} 
\\
          &       & NDCG  &       0.0012 &       \underline{0.0289} &       0.0135 &       0.0141 
&  \textbf{0.0621} 
\\
\cmidrule{2-8}          & \multirow{2}[2]{*}{Share} & Recall &       0.0074 &       \underline{0.1103} &       0.0074 &       0.0331 
&  \textbf{0.1324} 
\\
          &       & NDCG  &       0.0021 &       \underline{0.0470} &       0.0018 &       0.0133 
&  \textbf{0.0636} 
\\
\cmidrule{2-8}          & \multirow{2}[2]{*}{Follow} & Recall &       0.0153 &       \underline{0.0204} &       0.0102 &       \underline{0.0204} 
&  \textbf{0.0697} 
\\
          &       & NDCG  &       0.0048 &       \underline{0.0058} &       0.0028 &       0.0047 
&  \textbf{0.0380}
\\
\cmidrule{2-8}          & \multirow{2}[2]{*}{Mean} & Recall &       0.0389 &       \underline{0.0777} &       0.0490 &       0.0608 
&  \textbf{0.1217} 
\\
          &       & NDCG  &       0.0229 &       \underline{0.0408} &       0.0303 &       0.0355 
&  \textbf{0.0683} 
\\
    \bottomrule
    \end{tabular}%
    }
  \label{tab:single}%
\end{table}%

\subsection{Main Results (RQ1)}
In this subsection, we compare our proposed \model with state-of-the-art baseline models on the three experimental datasets. The comparison results with multi-behavior CF and CTR models are reported in Table \ref{tab:main_result}, and the comparison results with single-behavior CF models are illustrated in Table \ref{tab:single}. From the results, we can have the following observations:

\textbf{\model can achieve significant improvements across all types of behaviors over state-of-the-art methods.} From Table~\ref{tab:main_result}, we observe that \model achieves the highest Recall and NDCG performance across all types of behaviors than both current multi-behavior CF models and multi-behavior recommendation models, with mean NDCG improvements of 15.80\%, 7.49\%, and 35.74\% on Beibei, Taobao, and Tenrec respectively. Furthermore, from Table~\ref{tab:single}, we can find that \model can also generally obtain the best performance through simultaneous multi-behavior learning than single-behavior models with separate training for each behavior.
The results verify that \model can get better multi-behavior CF recommendation results.

\textbf{The multi-behavior CF models will negatively affect the performance on click behavior.} Comparing the multi-behavior CF models in Table~\ref{tab:main_result} and the traditional single-behavior CF models in Table~\ref{tab:single}, we can find that the multi-behavior CF models can obtain relatively better results in multiple behaviors recommendation, while they have a performance drop in the click behavior recommendation than the traditional single-behavior CF models. Single-behavior CF models only perform effectively on the click behavior, yielding suboptimal results on other behaviors because they lack training on such data.

\textbf{The multi-behavior CTR recommendation methods are unsuitable for CF tasks.} From the results, we can find that the state-of-the-art multi-behavior recommendation models in CTR scenarios have suboptimal performance in CF scenarios, even if they are equipped with the state-of-the-art LightGCN backbone. Therefore, it is meaningful to design a multi-behavior CF method with our proposed \pog and \model, which can achieve the best performance for multiple behavior recommendations simultaneously in the collaborative filtering setting. Therefore, it is crucial to propose a multi-behavior model in the CF scenario, which is the motivation of our proposed \model.

\subsection{Ablation Study (RQ2)}
In this section, we investigate the effectiveness of the designed components in our method with three model variants: (1) \textbf{w/o PO-all}, which remove all partial order components. We substitute the \pog with the interaction graph and treat all interactions with same weight. We also substitute POBPR with traditional BPR sampling. (2) \textbf{w/o POG}, which treats all nodes in the graph with equal weight during graph convolution. (3) \textbf{w/o POBPR}, which utilize traditional BPR sampling instead of POBPR.

\begin{figure}[tbp]
    \centering
    \includegraphics[width=\linewidth, trim=0cm 0cm 0cm 0cm,clip]{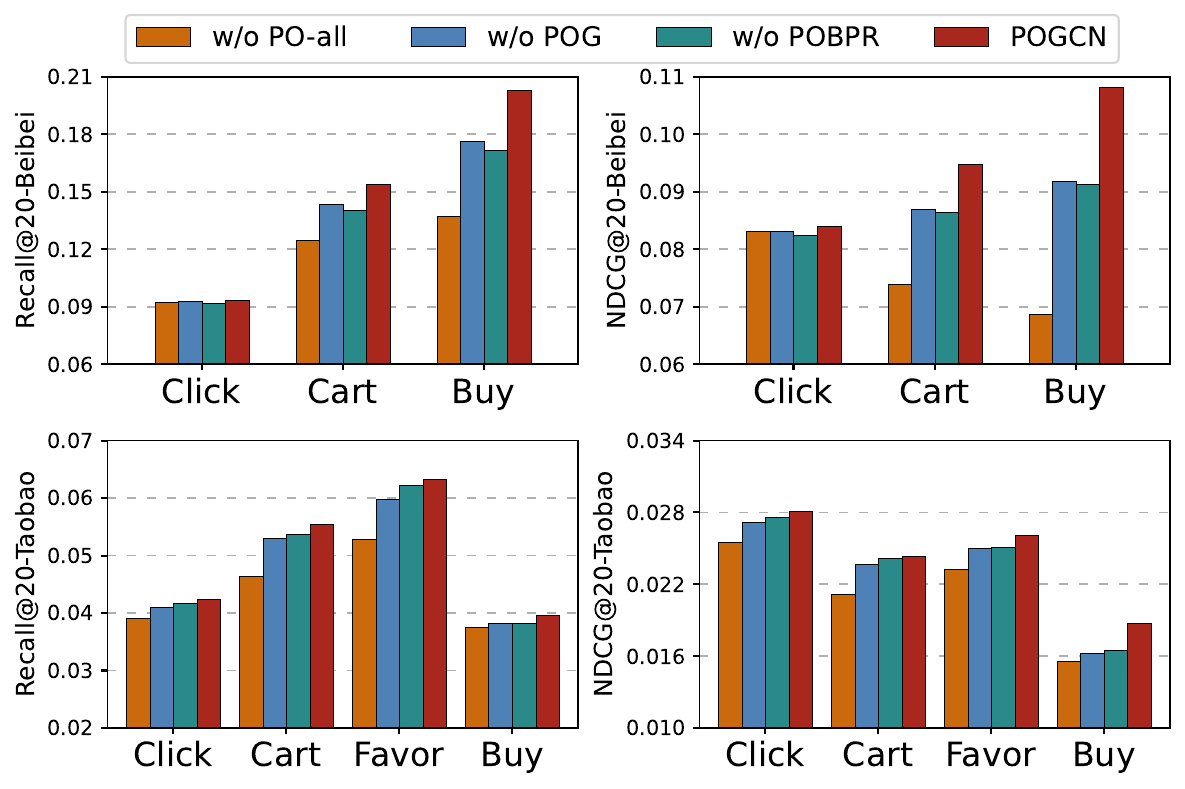}
    \caption{Ablation study results with three \model variants.}
    \label{fig:ablation}
\end{figure}

We perform ablation studies on Beibei and Taobao datasets, and the outcomes are presented in ~\autoref{fig:ablation}. Based on the figure, we make the following observations:

\textbf{Partial order relation is important for multi-behavior CF recommendation}. Substituting the \pog and POBPR with the interaction graph and traditional BPR sampling will significantly decrease the CF recommendation performance on all behaviors. This phenomenon verifies that Partial order relation provides a better description of multiple behaviors.

\textbf{POG and POBPR are necessary for multi-behavior CF recommendation}. From~\autoref{fig:ablation}, we observe that both POG and POBPR have a positive impact on the performance of multiple behaviors, indicating the importance of constructing customized convolution and training structures.


\subsection{Parameter Study (RQ3)}\label{sec:param_exper}
In this subsection, we aim to study the impact of key hyper-parameters $\tau$ and $r$ in \model.

\subsubsection{\textbf{Effect of \pog Parameter $\tau$}}
We investigate the effect of the  parameter $\tau$ of the importance degree between different behaviors with the range of [1.0, 5.0] with a step size of 1.0 for Beibei, and with the range of [0.2, 1.0] with a step size of 0.2 for Taobao.
As illustrated in the upper side of Figure \ref{fig:param_tau_r}, the best value of $\tau$ for Beibei is 3.0, while the best $\tau$ value for Taobao is 0.2. Therefore, the selection of parameter $\tau$ should be considered carefully for better \model performance.

\subsubsection{\textbf{Effect of Sampling Parameter $\gamma$}}
\begin{figure}[tbp]
    \centering
    \includegraphics[width=\linewidth, trim=0cm 0cm 0cm 0cm,clip]{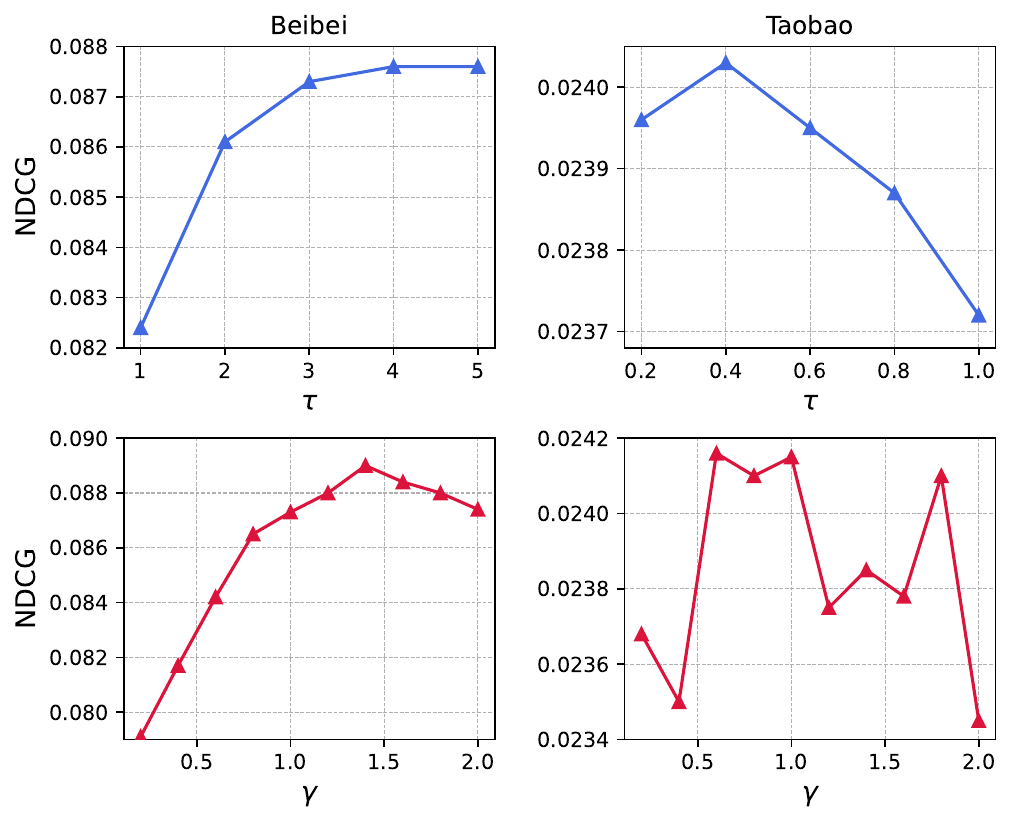}
    \caption{Parameter study of $\tau$ and $\gamma$ on Beibei and Taobao over the Mean NDCG metric.}
    \label{fig:param_tau_r}
\end{figure}

We also evaluate the impact of different partial order sampling values of $\gamma$, with the range of [0.2, 2.0] with a step size of 0.2.
From the second row of Figure \ref{fig:param_tau_r}, we can find that the too-small $\gamma$ value will lead to too similar between different behaviors and lead to suboptimal performance.
The selection of the $\gamma$ value is contingent upon the specific characteristics of each dataset, necessitating a careful choice to optimize performance. For instance, an $\gamma$ value of 1.4 is well-suited for the Beibei dataset, whereas a lower value of 0.6 is preferable for the Taobao dataset.

\begin{table}[tbp]
  \centering
  \caption{Results of online A/B tests in the industrial platform.}
  \resizebox{0.975\linewidth}{!}{
    \begin{tabular}{l|ccccc}
    \toprule
    A/B Test & PCTR & UCTR & CVR & GMV & StayTime\\
    \midrule
    v.s. LightGCN & +2.83\% & +2.02\% & +1.69\% & +2.84\% & +1.62\% \\
    \bottomrule
    \end{tabular}%
   }
  \label{tab:abtest}%
\end{table}%

\subsection{Online Evaluation (RQ4)}
We conducted an online A/B test on the homepage of Alibaba. In this experiment, our model served as a recall model, replacing the existing online best-performed graph-based recall model---LightGCN. In addition to clicking behavior, POGCN also considers three other valuable behaviors: ``adding to cart'', ``favoring'', and ``buying''. Table~\ref{tab:abtest} presents the average relative performance variation over a successive month for about 1 billion users and 0.2 billion items. 

From the results, we observe that \model shows performance improvements of $+1.69\%$ in CVR, $+2.84\%$ in GMV, and $+1.62\%$ in StayTime compared to LightGCN. This indicates that considering deep behaviors such as ``adding to cart'', ``favor'', and ``buy'' can enhance the interaction depth and recommendation conversion, thereby increasing the intent of purchasing and viewing more items. Besides, equipped with POGCN, the recommender system is able to provide more accurate recommendations to users, leading to a $+2.83\%$ increase in PCTR and $+2.02\%$ improvement in UCTR. \model achieves this through the simultaneous optimization of multiple types of behaviors, thereby enhancing user willingness to click more recommended items.

Therefore, all the A/B testing results validate that $\pog$ and the equipped $\model$ are more suitable than the state-of-the-art online graph collaborative filtering models.

\section{RELATED WORKS}



\subsection{Graph Collaborative Filtering}
In recent years, graph neural networks (GNNs) have achieved superior performance on a wide range of relation modeling tasks~\cite{LAGCN,zhang2023integrating,wang2023pattern,wang2024towards}, such as link prediction~\cite{bei2023cpdg,zhang2018link,wu2020surveyGNN,tan2023collaborative}, node classification~\cite{kipf2016semigcn,NEGCN,rong2020dropedge,NIPS2017_graphsage}, and anomaly detection~\cite{bei2023reinforcement,xiao2023imputation,xiao2023counterfactual,zhang2022contrastive}. 

As a mainstream link prediction task, there have been extensive works for graph collaborative filtering based on the power of GNNs~\cite{chen2022differential,gao2023survey}. Representatively, NGCF~\cite{wang2019ngcf} employs a GNN propagation mechanism that bridges users to items and users to users, extracting graph embeddings for each entity. Moreover, LightGCN~\cite{he2020lightgcn} simplifies the approach by omitting the non-linear parametered operators of NGCF and instead, aggregates the layer-wise embeddings through a weighted summation.
Recent studies have further incorporated data augmentation and self-supervised learning techniques to bolster the performance of graph collaborative filtering models~\cite{gao2023survey,yu2023self}. Notably, SGL~\cite{wu2021sgl} pioneers the integration of self-supervised learning on the user-item graph by generating diverse node perspectives. It leverages contrastive learning to align these multiple views of identical nodes, while simultaneously reducing the alignment with nodes that are distinct. NCL~\cite{lin2022ncl} further refines the neighbor set by incorporating semantic neighbors, guided by the principles of contrastive learning. Meanwhile, SimGCL~\cite{yu2022simgcl} offers a streamlined yet potent graph contrastive learning approach through the strategic elimination of extraneous augmentations. However, these models mainly focus on the scenario of a single type of interaction behavior, without the consideration of how to use multiple interaction behaviors for the recommendation.

\subsection{Multi-Behavior Collaborative Filtering}
Collaborative filtering recommendation was originally designed for single-behavior recommendations~\cite{su2009cfsurvey, koren2021advancescfsurvey,huang2024large}. Examples of such recommendations include sequence-based~\cite{zhou2018din,huang2023ALDI,MPAD,chen2022generative} and graph-based~\cite{lin2024box,he2020lightgcn,wang2019ngcf,zhou2021temporal}. However, in real-world scenarios, users and items exhibit multiple interaction behaviors~\cite{huang2021mbsurvy}, such as clicks, purchases, or ratings, instead of focusing solely on a single behavior like clicks~\cite{jin2020mbgcn,cheng2023mbcgcn,zhou2022multi}.

Recent efforts for multi-behavior collaborative filtering mainly focus on utilizing information from multiple types of behaviors for user modeling to enhance recommendation performance for sparse target behaviors, such as using data from clicks, carts, and purchases to model interactions in the purchase behavior domain~\cite{wang2023mtlsurvey,wei2022cml,xia2020matn}.
Representatively, NMTR~\cite{gao2019nmtr} incorporates a multi-task learning framework that acknowledges the cascading relationship among different user behaviors, allowing for more accurate modeling of user preferences based on a comprehensive set of interactions for target behavior recommendations.
Recently, MB-CGCN~\cite{cheng2023mbcgcn} further employs a sequence of GCN blocks that correspond to different user behaviors to capture the complex dependencies between different user behaviors like views, clicks, and purchases. It enhances recommendation performance by exploiting the cascading nature of these behaviors, where each behavior's embeddings are transformed and used as input for the next behavior's embedding learning.
Nevertheless, these studies mainly focus on utilizing all types of behaviors to enhance the target behavior performance, while the simultaneous optimization of multiple behaviors has been highly neglected, resulting in poor performance for non-target behaviors.

\subsection{Multi-Task Recommendation}
Multi-task recommendation aims to effectively model relationships between different recommendation tasks in a multi-task learning framework~\cite{liu2023deep,yan2023mbccrgcn,zhang2021MTLsurvey,crawshaw2020MTLDeepsurvey}.
These models have achieved state-of-the-art performance in the ranking and prediction stage of recommender systems, such as simultaneously predicting the click-through rate (CTR) and conversion rate (CVR)~\cite{bei2023nrcgi,ma2018essm,su2024stem}.

The most common method of multi-task learning is Shared Bottom~\cite{caruana1997multitask}, which uses the coupled input to predict each task individually. 
ESMM~\cite{ma2018essm} utilizes a novel structure that models CVR over the entire space of impressions by employing auxiliary tasks of predicting click-through rate and click-through\&conversion rate.
MoE~\cite{jacobs199moe}, MMoE~\cite{ma2018mmoe} utilizes a mixture of experts architecture, where each "expert" is a specialized network in a shared structure, and multiple gating networks then learn to weigh the contribution of each expert for a given task.
Further, PLE~\cite{tang2020ple} separates shared and task-specific components, employing multi-level experts and gating networks, and introducing a novel progressive separation routing mechanism. This allows PLE to extract deeper, more relevant knowledge for each specific task while mitigating harmful interference between tasks.
However, these models are mainly designed for the prediction stage of recommendations. As the previous analyses, they lack the capability to be adopted in the collaborative filtering stage.

\section{CONCLUSION}


In this paper, we study the seesaw recommendation problem in current multi-behavior CF models.
To this end, we introduce the \textit{Partial Order Recommendation Graphs (POG)} and \textit{Partial Order Graph Convolutional Networks (POGCN)}, which has significantly advanced the field of multi-behavior collaborative filtering for recommender systems. POGCN offers an update in the infrastructure of multi-behavior CF tasks and presents a groundbreaking solution to the longstanding issue of effectively representing diverse user interactions within a single CF framework. Demonstrating superior performance in both offline experiments on three benchmark datasets and online A/B tests on the industrial system, and practically serving over a billion users in a major shopping platform, POGCN not only elevates the capability of recommendation research but also paves the way for future research and optimizations in the realm of large-scale multi-behavior recommender systems.

\begin{acks}
This work was supported in part by the National Natural Science Foundation of China (Grant No. 62272200, U22A2095, 61932010).
\end{acks}


\bibliographystyle{ACM-Reference-Format}
\bibliography{reference}

\appendix
\section{Theoretical Details}
\subsection{Completeness Proof of Partial Order of Behavior Combination}
\label{sec:comp_combination proof}

\begin{proof}
We aim to construct a rank function that maps the subsets partitioned from the set \(\gC\) by the binary relation \(\leq_c\) to a positive integer, thereby establishing \((\gC, \leq_c)\) as a graded partial order set. Define \(\gC_i <_c \gC_j\) if \(\gC_i \leq_c \gC_j\) and \(\gC_i \neq \gC_j\).

Consider an arbitrary element \(\gC_i\) from \(\gC\). Using the binary relation, we partition \(\gC\) into three subsets:
\begin{itemize}
    \item \(L_c\), containing all \(\gC_j\) such that \(\gC_j <_c \gC_i\),
    \item \(E_c\), containing all \(\gC_j\) such that \(\gC_j = \gC_i\) or \(\gC_i\) and \(\gC_j\) are incomparable,
    \item \(G_c\), containing all \(\gC_j\) such that \(\gC_i <_c \gC_j\).
\end{itemize}

According to Definition \ref{def3}, for any \(\gC_{j_1} \in L_c\), \(\gC_{j_2} \in E_c\), and \(\gC_{j_3} \in G_c\), it holds that \(\gC_{j_1} <_c \gC_{j_2} <_c \gC_{j_3}\). Elements in \(E_c\) are considered of equal significance.

This partitioning process is then recursively applied to \(L_c\) and \(G_c\) until these sets are empty. Ultimately, we obtain the final partition: \(E_{c_1}, E_{c_2}, \ldots, E_{c_n}\), satisfying \(\gC_{j_1} <_c \gC_{j_2} <_c \cdots <_c \gC_{j_n}\) for any \(\gC_{j_1} \in E_{c_1}\), \(\gC_{j_2} \in E_{c_2}\), \ldots, \(\gC_{j_n} \in E_{c_n}\).

Define \(\rho_c(\gC_{j_k}) = k\) for all \(\gC_{j_k} \in E_{c_k}\) and \(k = 1, 2, \ldots, n\). It is evident that \((\gC, \leq_c)\) equipped with \(\rho_c\) satisfies Definition \ref{def2}.
\end{proof}

\subsection{Equivalence Explanation of POBPR} \label{sec:po-BPR}
Here, we will explain that such a distribution change is equivalent to transforming the coefficients in the case where behavior combinations are treated as multi-tasks. We have the likelihood function, following BPR~\cite{rendle2009bpr}:
\begin{equation*}
    L(\theta) = \prod_{h=1}^{H} \prod_{(u,i, j)\in \gD_h^{+} \times \mathcal{I}} p(i >_u j | \theta), 
\end{equation*}
where $\gD_h^{+} = \{(u,i)|\mB_{ui}=\gC_h\}$ represent the interaction set of behavior combination $\gC_h$. For the sake of following discussion, we let $p(i >_u j|\theta) := \sigma(\hat{Y}_{ui}(\theta)-\hat{Y}_{uj}(\theta))$, and can get the partial order BPR loss function:
\begin{equation*}
    \begin{split}
    \mathcal{L}_{\text{POBPR}} &= \ln p(\Theta|>_u ) \\
    &= \ln p(>_u |\Theta)p(\Theta) \\
    &= \ln \prod_{h=1}^{H} \prod_{(u,i,j)\in \gD_h^{+} \times \mathcal{I}} p(i >_u j | \theta) + \ln p(\theta) \\
    &= \sum_{h=1}^{H} \sum_{(u,i,j)\in \gD_h^{+} \times \mathcal{I}} \ln \sigma(\hat{Y}_{ui}(\theta)-\hat{Y}_{uj}(\theta)) + \ln p(\theta) \\
    &= \sum_{h=1}^{H} \alpha_h \mathcal{L}_{\text{BPR}, \gC_h} + \ln p(\theta), \\
    \end{split}
\end{equation*}
where $\alpha_h \propto P(\gC_h)$ is the coefficient of the behavior combination $\gC_h$ task. So when we change $\gamma$, the coefficients in the multi-task loss function will also change equivalently.

\section{Experimental details}
\subsection{Dataset Details}\label{sec:data_appendix}
We adopt three publicly available datasets for offline evaluation. The detailed description of the datasets is as follows:
\begin{itemize}
    \item \textit{Beibei}\footnote{\url{https://github.com/Sunscreen123/Beibei-dataset}}~\cite{gao2019nmtr} is gathered from Beibei platform, one of China's premier e-commerce platforms specializing in baby products. It encompasses the interactions of 21,716 users and 7,977 items, with three types of user-item behaviors: click, cart, and buy.
    \item \textit{Taobao}\footnote{\url{https://tianchi.aliyun.com/dataset/dataDetail?dataId=649}}~\cite{zhu2018taobao} is a wide used multi-behavior dataset provided by Alibaba, one of the biggest e-commerce platforms in China. It contains the activities (including clicks, carts, favors, and buys) of 26,213 users toward 64,822 items between Nov. 2017 and Dec. 2017.
    \item \textit{Tenrec}\footnote{\url{https://static.qblv.qq.com/qblv/h5/algo-frontend/tenrec_dataset.html}}~\cite{yuan2022tenrec} is a content recommendation dataset collected from two different feeds recommendation apps of Tencent. We utilize the video scenario subset for experiments, with 19,035 users, 15,539 items, and four types of user-item behaviors: click, like, share, and follow.
\end{itemize}


\subsection{Baseline Details}\label{sec:baseline_appendix}
We compare our proposed \model with thirteen representative state-of-the-art models into three main categories as follows:

\noindent (I) \textbf{Multi-behavior CF models}: 
\begin{itemize}
\item \textit{MC-BPR}~\cite{loni2016mcbpr} utilizes multi-behavior weight to modulate the importance of different behaviors in the BPR loss function.
\item \textit{NMTR}~\cite{gao2019nmtr} is an expansion of NCF~\cite{he2017ncf} in multi-behavior recommendation, adhering to the cascade rule.
\item \textit{GHCF}~\cite{chen2021ghcf} is a graph-based approach for enhancing target behavior recommendation through multi-task learning.
\item \textit{MB-GMN}~\cite{xia2021mbgmn} is a graph meta-learning based CF model to improve target behavior recommendations.
\item \textit{MB-CGCN}~\cite{cheng2023mbcgcn} combines multi-behavior cascade rule to enhance graph convolution networks for the target behavior recommendation.
\item \textit{IMGCF}~\cite{zhang2023imgcf} employs a multi-task learning paradigm for collaborative filtering on multi-behavior graphs, enhancing sparse behavior learning by leveraging information from behaviors with ample data.
\end{itemize}

\noindent (II) \textbf{Multi-task recommendation models}: 
\begin{itemize}
\item \textit{ESMM}~\cite{ma2018essm} employs a feature representation transfer learning strategy for multi-task CTR recommendations.
\item \textit{MMoE}~\cite{ma2018mmoe} adapts the MoE structure to multi-task learning by sharing the expert submodels across all tasks, while also having a gating network trained to optimize each task.
\item \textit{PLE}~\cite{tang2020ple} is a progressive layered extraction model with a novel sharing structure design for multi-task CTR recommendations.
\end{itemize}

\noindent (III) \textbf{Single-behavior CF models}: 
\begin{itemize}
\item \textit{MF-BPR}~\cite{rendle2009bpr} is a widely used matrix factorization strategy with the assumption that the positive items should score higher than negative ones.
\item \textit{NCF}~\cite{he2017ncf} is a CF model that leverages a multi-layer perceptron to learn the user-item interaction function.
\item \textit{NGCF}~\cite{wang2019ngcf} explicitly encodes the collaborative signal in the form of high-order connectivities by performing graph embedding propagation.
\item \textit{LightGCN}~\cite{he2020lightgcn} is a simplified variant of NGCF by including only the most essential components.
\end{itemize}

\end{document}